\documentclass[a4paper]{ESASPStyle}
\usepackage{epsfig}

\begin{document}
\newcommand{\gras}[1]{\mbox{\boldmath $#1$}}
\def\msun{$\rm M_\odot~$}
\def\lsun{$\rm L_\odot$}
\def\teff{$T_{\rm eff}$}
\def\Mdot{$\dot{M}$}
\def\mbol{$M_{\rm bol}$}
\def\tbce{$T_{\rm bce}$}
\def\simgt{\lower.5ex\hbox{$\; \buildrel > \over \sim \;$}}
\def\simlt{\lower.5ex\hbox{$\; \buildrel < \over \sim \;$}}
\def\ltsima{$\; \buildrel < \over \sim \;$}
\def\gtsima{$\; \buildrel > \over \sim \;$}

\title{Models of Stellar structure for asteroseismology}

\author{F. D'Antona\inst{1}, J. Montalb\'an\inst{2} \&
I. Mazzitelli\inst{3}}

  \institute{INAF, Osservatorio di Roma, via di Frascati 33, I-00040
Monteporzio, Italy \and Institut d'Astrophysique, Liege, Belgium
  \and IASF--CNR, Roma, Italy
    }

\maketitle 

\begin{abstract}
Among the problems still open in the study of stellar structure, we discuss
in particular some issues related to the study of convection.
We have recently built up complete stellar models, adopting a consistent
formulation of convection both in the non gray atmosphere and in the
interior, to be used for non adiabatic pulsational analysis, and discuss in
some details two problems which have been clarified by these models: the
physical interpretation of one of the main sequence ``B\"ohm-Vitense gaps",
and the necessity of parametrizing pre main sequence convection differently
from the main sequence convection.
We also report preliminary results of the application to the solar model of
the non--local turbulence equations by \cite*{canuto-dubovikov1998} in the
down--gradient approximation.
 \end{abstract}

\section{Introduction}
The list of unsolved problems in stellar structure is long: microscopic
diffusion; mass loss; rotation, its evolution and its effects on chemical
mixing; magnetic fields, and their interaction with rotation and convection;
and an adequate model of convection able to describe the overadiabaticity in
the shallow layers, and non-local aspects like overshooting.
Asteroseismology, and especially asteroseismology from the space, is destined
to be very useful in providing us with additional constraints to our models.
Here, we only touch the problem of convection in stars, and in particular the
problem of envelope convection. The convective envelope in low mass stars
has a fundamental role in $\delta$Scuti, $\gamma$Doradus and solar-type
oscillations. Furthermore, the boundaries between the corresponding domains
of instability in the HR diagram seem to be linked to changes in the
convection features. The non adiabatic seismological analysis requires
knowledge of the entire stellar structure up to the atmosphere, and
selfconsistent convection models for the atmosphere and interior integration
are necessary for this use. On the other hand, progress in the study of
convection is slow enough that today's main attitude with respect to this
problem is the following: parametrize in a rough way the overadiabatic
convection, by using constraints either coming from the analysis of stellar
spectra or from other structural observational information (e.g. the p--modes
solar patterns, or the depletion of light elements in the stellar envelopes,
or empirical radii constraints). This can be sufficient for studying solar
oscillations, but it is difficult to extend models so much parametric to
regions of the HR diagram in which the observational constraints are scarce
(or none). We have been attempting to build up complete stellar models to be
used for non adiabatic pulsational analysis. These models have clarified some
interesting problems in stellar evolution which we discuss here in more
detail: {\it i)}: they have provided an interpretation for the ``B\"ohm
Vitense gap" observed in open clusters at B--V$\sim$0.35
(\cite{rachford-canterna2000}); {\it ii)}: they have made explicitly clear
the necessity of parametrizing pre main sequence convection differently from
the main sequence convection, possibly implying that there is an hidden
``second parameter" acting to change the convection behaviour during the
first phases of stellar evolution.
Finally, we show the results of preliminary computation of non local
convection in the solar model, which indicate that the overshooting expected
at the bottom of the solar convective envelope is $\sim 0.02$H$_P$.

\section{Convection in complex evolutionary phases}
A good example of complex models of stellar structure are the
intermediate mass Asymptotic Giant Branch (AGB) stars, those which evolve
through the phase called Hot Bottom Burning (HBB).
These stars are extremely important for the chemical evolution of galaxies,
and they are probably the key ingredient to develop primordial chemical
inhomogeneities among Globular Clusters stars. During 90\% of their life,
they are fueled by hydrogen, whose stationary burning in an external shell
is interrupted by the sudden ignition of the helium beneath. During the
ensuing thermal runaway (the Thermal Pulse --TP-- phase), the hydrogen
envelope expands, hydrogen burning stops and external convection reaches into
the helium nuclearly processed layer, provoking the `third dredge up' phase
by which nuclearly processed material appears  at the stellar surface. In the
most massive AGB stars (4--8\msun), the stationary hydrogen burning shell is
partially contained into the convective envelope, that is, the bottom of the
convective envelope attains temperatures so large that nuclear reactions
take place there. The CN --or even CNO, for low metallicity stars
(\cite{ventura2001})-- nuclearly processed material is convected to the
surface and then convected back to the bottom. So the nuclear processing
which can be observed in the atmospheres of these stars is due to the
combined action of HBB and of the third dredge up. In spite of the
fundamental importance of these objects evolution, for sure this phase is
not easily modeled! One of the big problems is the (unknown) mass loss, which
affects the yield of the processed chemistry both through the time dependent
loss of the envelope into the interstellar medium, and through altering the
temperature structure of the envelope. In addition, these stellar models are
heavily dependent on the way we model convection it {\it all its aspects}:

\begin{enumerate}

\item  {\it the convective model}, that is the convective flux and
temperature gradients computation;

\item  {\it overshooting}, that is mixing outside the formal convection
boundaries. For AGBs, this includes both the problem of core--overshooting,
which affects the relation between initial mass and mass at the beginning of
the TP phase, and the possible overshooting below the convective envelope,
which affects the nucleosynthesis;

\item {\it time dependence}: mixing must be treated as non instantaneous, and
coupled with nucleosynthesis, for all the elements for which the convective
mixing timescale is of the same order of the nuclear burning timescale.

\end{enumerate}

\section{Convection in more ``normal" stars}    
In the present discussion, we leave entirely aside the ``big problems" of
convection modelling ---apart from the last section. While turbulence in
stars is compressible and non local, we must accept that for a long time, for
general purposes of computing stellar structure for any mass, chemistry and
evolutionary phase, we still will deal with incompressible and local models.
3D radiation hydrodynamics (RHD) simulations still miss the computer power
needed to deal with deep envelope convection, although great insight has been
obtained in the atmospheric studies. Interesting results are available by 2D
simulations. Analytic non--local convection models have recently been applied
to stellar atmospheres of A stars (\cite{kumo2002}) and we may foresee
important developments also in this area. Among the local models, the Mixing
Length Theory (MLT) by \cite*{bv1958} certainly is still dominant, in spite
of its limitations. Other local models have become available, among them the
Full Spectrum Turbulence (FST) model by \cite*{cm1991} and its variant in
\cite*{cgm1996}, which have been adopted for a variety of applications in the
latest 10 years.

\begin{figure}[!bt]
  \begin{center}
    \epsfig{file=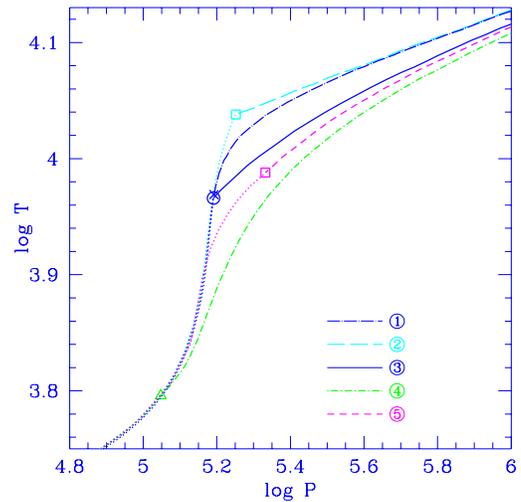, width=7cm}
  \end{center}
\caption{Stratification of temperature vs. pressure in the atmosphere and
sub--atmosphere of solar models by Montalb\'an et al. (2003).
Track 1 is the model employing FST convection both in the atmosphere and
interior, the match is done at $\tau_{\rm ph}=10$; models from 2 to 4 have
MLT atmospheres with $\alpha_{\rm atm}=0.5$, and MLT interiors with
$\alpha_{\rm in}$\ chosen so to fit the solar radius. The differences are due
to the different choices of matching points: 2: $\tau_{\rm ph}=100$\ and
$\alpha_{\rm in}=6.3$; 3: $tau_{\rm ph}=10$\ and $\alpha_{\rm in}=2.3$; 4:
$\tau_{\rm ph}=1$\ and $\alpha_{\rm in}=1.75$. Model 5 employes the AH97
atmospheres down to $\tau_{\rm ph}=100$, and has $\alpha_{\rm in}=1.9$ (the
model shown in Figure \ref{fig1}). We see that the subatmospheric structure
is very different for the different models, in spite of the `solar radius'
calibration. \label{fig2}} \end{figure}

As no available theory is self consistent, the dominant attitude in
stellar studies --unless there are particular requirements-- is
to use the convective model as a `black box' simply to infer the stellar
properties in the region in which convection becomes adiabatic.
In general, in fact, the main important property which a convection
model is required to provide is the stellar radius: whatever is in the `black
box' of convection, we first of all wish to know the radius (or the \teff,
or, equivalently, the entropy jump between the adiabatic interior and the
surface). The solar radius in particular is generally adjusted in the models
by varying the convective efficiency. In the MLT this can be done by varying
the `mixing length' parameter, that is the ratio of mixing length to pressure
scale height $\alpha=l/H_p$.
Recently it has been stressed the importance of surpassing the Eddington
approximation, or the gray atmosphere approximations to correctly describe
the optical atmosphere. For many stellar situations, a non gray model
atmosphere ---which also must reproduce the observed spectral features---
includes as necessary ingredient a treatment of convection, which, in the
MLT, will be characterized by a given $\alpha_{\rm atm}$\ down to the
optical depth $\tau_{\rm ph}$\ at which the atmosphere will be matched to the
interior computation. In the interior, convection will be characterized by a
given $\alpha_{\rm in}$. Thus, actually, a `modern' MLT structure will have
three parameters: $\alpha_{\rm atm}$, $\alpha_{\rm in}$\ and $\tau_{\rm ph}$.
Figure~\ref{fig9} illustrates this classic problem by using our recent models
(\cite{mdkh2003}) which adopt as boundary conditions the NextGen atmospheric
grid (\cite{AH97}, hereinafter referred to as AH97), computed by assuming
MLT convection with $\alpha_{\rm atm}=1$, down to optical depth $\tau_{\rm
ph}=100$.

Adopting $\alpha_{\rm in}=1.9$\ for the interior computations, we obtain a
set of tracks, among which the solar track passes through the solar
location at the solar age. The solar mass track with $\alpha_{\rm in}=1.0$,
on the contrary, is $\sim 380$K cooler at the solar luminosity. This well
known fact remembers how large is the variation in \teff\ allowed by changing
the MLT parameter. The meaning of this solar radius adjustment is simply the
following: when using the MLT, {\it the entropy jump, necessary to fit the
solar $T_{eff}$,} between the adiabatic layers and the surface corresponds
to what is obtained with the MLT and $\alpha=1.9$\ up to $\tau_{\rm ph}=100$,
and with MLT and $\alpha=1$\ up to the surface. Of course, the model can not
tell us anything about the temperature gradient {\it layer by layer} within
the envelope. If we only wish to use convection as a `black box', we just
take the $\alpha_{\rm in}=1.9$\ model for the Sun\footnote{and generally we
forget, or at least do not even mention, that $\alpha$\ is actually
$\alpha_{\rm atm}=1$ for most of the overadiabatic region, which is contained
in the model atmosphere grid.}. Of course, our assumption will only be valid
for the Sun itself.

An interesting broadening of of this point of view to a wider part of the HR
diagram has been given by \cite*{ludwig99}, who performed detailed 2D
numerical RHD calculations of time-dependent compressible convection, in the
range 4300$< T_{\rm eff} <$ 7100K, 2.54$< \log g <$4.74 (cgs) for solar
composition. They used these models to `calibrate' the effective $\alpha$\
for these envelopes, that is the value of $\alpha$\ which provides the same
specific entropy jump (using a gray atmosphere, therefore this approximation
defines a unique average $\alpha$\ parameter) than their 2D models. They
found values from 1.3H$_{\rm p}$\ for F dwarfs up to 1.8H$_{\rm p}$ for K
giants. This calibration of $\alpha$, again, does not tell us anything about
the temperature gradient layer by layer.

The approach of keeping convection as an entirely black box is very useful,
but it has --obviously-- many limitations. In particular, if we wish to study
the {\it excitation} of oscillation instabilities in stars we must know the
full stratification of physical quantities in the star, up to the atmosphere.

\section{Convection modelling in full stellar models}
We have to recognize that no general purpose convection model is presently
available, which can be meaningfully applied to any stellar structure in
order to know the layer by layer stratification of physical quantities.
Nevertheless, recently there have been several attempts to produce {\it
entire} stellar models, which satisfy some observational constraints on the
atmospheric and envelope convection, and in which the atmospheric and the
envelope convection are matched `smoothly' each other. A smooth match is at
least technically necessary for stellar stability studies, to avoid
discontinuities in the physical
quantities. A prototype
of these models has been discussed by \cite*{schlattl1997} for the Sun. They
built up a non gray 1D model atmosphere with $\alpha_{\rm atm}=0.5$\ down to
$\tau_{\rm ph}=20$, and matched it to a subatmosphere and MLT interior,
guided by the results of 2D hydro models. To do this, they had to vary the
$\alpha_{\rm in}$\ parameter in the interior computation, with the aim to
provide the solar radius and obtain a temperature stratification similar to
that of the 2D models. The aim of this study was to explore the influence of
the physical inputs on the solar p--modes.

Of course, for the Sun we have an enormous number of constraints ---from the
p--modes themselves, and from the precise knowledge of the solar radius---
which help to build up a fully parametric model. But what should we do to
extend the analysis to other stars?
What kind of model atmosphere can we use, and how do we produce a
`meaningful' match of the atmospheric and interior computation? There are not
yet many model atmosphere grids available, and many of them have been
computed for ``atmospheric purposes" only, that is to produce an adequate
modelling of the region from which the stellar spectum emerges, so they are
not entirely apt to be used as boundary conditions for full stellar models.
For instance, the quoted AH97 models, computed with the PHOENIX code, adopt
MLT convection with $\alpha=1$\ and a total of 50 layers down to $\tau=100$.
\cite*{heiter2002} have adopted a new version of Kurucz's (\cite{ku93}) model
atmospheres (the Vienna--ATLAS9 code), in which the convection model is
either the MLT with $\alpha=0.5$\ or the FST (both in the \cite{cm1991} and
in the \cite{cgm1996} versions), and they have increased the number of layers
in the latter models to 288. We have recently used all these grids of model
atmospheres to explore the meaning of the match between atmosphere and
interior, and what happens when one extends to other models (pre main
sequence, and giants) the approximations adopted for the solar model. The
extensive results of this study are presented elsewhere (\cite{mdkh2003}).
Here we show, in Figure \ref{fig2}, that the different solar models which can
be built up, all satisfying the constraint of the solar fit, can have very
different subatmospheric structures, depending on the choice of the
convection efficiency in the different layers.

In Figure \ref{fig2} we also see that the model adopting the FST convection
both in the atmosphere and in the interior does not show discontinuities in
the T vs. P stratification. The structure smoothly passes from a very
inefficient convection in the atmosphere, similar to that of the MLT models
with $\alpha=0.5$, to such an efficient convection inside, that the
precise solar fit is easily obtained (see e.g.
\cite{cm1991})\footnote{Remember that, whatever the choice of the free
parameters in the FST convection, it is {\it not} possible to obtain a 1\msun
location, at L$_\odot$, with \teff\ much smaller or larger than the solar
\teff, contrary to the MLT models.}.
\begin{figure}[!ht]
  \begin{center}
    \epsfig{file=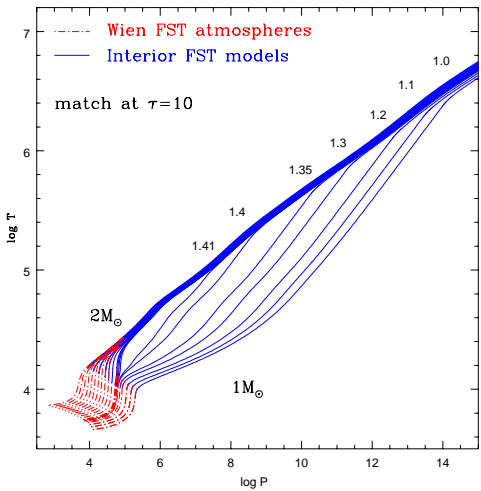, width=7cm}
  \end{center}
\caption{Temperature vs. pressure stratification in models from 2 to
1\msun\ at $10^8$yr, built with the Vienna-ATLAS9 FST boundary conditions.
Both the atmosphere (dot-dashed) and the interior (full lines) are shown.
The transition between almost radiative envelope models (down to 1.42\msun)
and models with well developed envelope convection (below 1.42\msun) is
clearly shown. The same transition is much smoother in models with MLT
convection, having also in atmosphere $\alpha$ larger than 1.\label{fig3}}
\end{figure}
 \begin{figure}[ht] \begin{center}
  \epsfig{file= 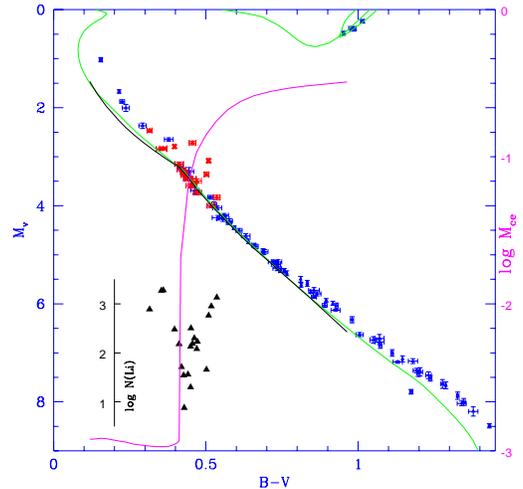, width=7cm}
 \end{center}
\caption{HR diagram of the Hyades from Hipparcos data
(Debrujine et al. 2000, 2001), and an isochrone of
600Myr built by the present models (Montalb\'an et al. 2003) compared with an
isochrone built by gray models (Ventura et al. 1998a). MLT models with
$\alpha=1.6$\ which fits the Solar location have a smoother transition
(Mazzitelli \& D'Antona 1993). We also show the `lithium dip'
(Boesgaard \& Tripicco 1986) (triangles), and the logarithm of the
convective mass (scale to the right).}
\end{figure}
Computing ``full FST" structures, then, has as a first a ``formal" advantage,
as it helps to avoid the problem of discontinuities in the physical
quantities for instability studies. In addition, there are other, more
physical, reasons to use this model as probe:  a series of works
on {\it 1)} \teff\ determination from H$_\alpha$\ and H$_\beta$\ lines (e.g.
\cite{fuhrmann1993}; \cite{vv1996});
{\it 2)}:
theoretical predictions of H$_\alpha$\ and H$_\beta$\ from 1D models
(\cite{gardiner1999}) and from 2D models (\cite{steffen-ludwig1999}); {\it
3)}: theoretical predictions of $b-y$\ and $c$\ Str\"omgren indices
(\cite{smalley-kupka1997}) and abundance determinations (\cite{heiter1998})
indicate that, even if a 1D, homogeneous model can not explain all the
spectroscopic and photometric observations, model atmospheres which predict
temperature gradients closer to the radiative gradient are in better {\it
overall} agreement with the observations. These arguments favors those models
in which convection in the atmosphere is less efficient than predicted by
models having $\alpha>1$), and led \cite*{heiter2002} to adopt either the FST
model or the $\alpha=0.5$\ MLT models to compute their atmospheric grids.
These were also further motivations to produce ``full FST" stellar models by
use of their atmospheric grids.

\begin{figure}[!bt]
  \begin{center}
    \epsfig{file= 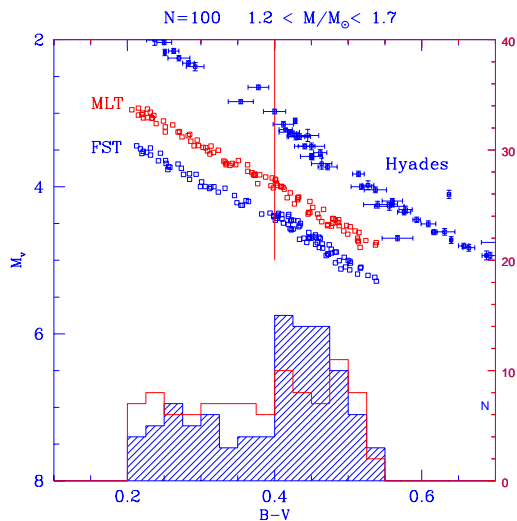, width=7cm}
  \end{center}
\caption{The Hyades Hipparcos HR diagram is compared with two
simulations containing 100 stars between 1.2 and 1.7\msun, randomly
extracted from a Salpeter mass function distribution, following the FST or
MLT mass--color isochrone of 600~Myr. The MLT simulation is shifted by
+0.7~mag, and the FST one by 1.2~mag for clarity. At the bottom, the FST
(dashed) and MLT color histograms are shown (scale at the right). The FST
mass--color distribution produces a `gap' at B--V$\sim$0.35, while the MLT
isochrone does not. The good correspondence with the Hyades gap indicates
that the transition between structures which are convective only in the
surface layers, and structures which show a well developped convection also
in the interior is {\it very sharp}\label{fig5}} \end{figure}
\begin{figure}[!ht]
  \begin{center}
    \epsfig{file= 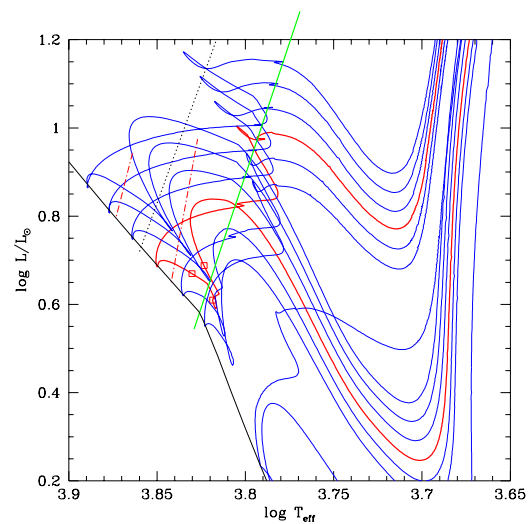, width=7cm}
  \end{center}
\caption{The general HR diagram of the computed tracks, which extend from the
PMS to the MS and then to the Giant Branch. The tracks
of 1.2, 1.4, 1.45, 1.5, 1.55, 1.6, 1.65\msun are shown. The diagonal (green)
line on the right shows schematically the separation line between higly
convective structures on the right and structures with convection limited to
the atmosphere (on the left). The diagonal dotted (black) line indicates the
observational red edge of the $\delta$ Scuti instability strip, according to
Pamyatnykh (2000). The two dash--dotted (red) lines indicate the
observational boundaries of the $\gamma$ Dor strip (Zerbi 2001). The MS at
10$^8$yr age is also shown.\label{fig6}} \end{figure}

We may ask whether the ``full FST" models have shown features interesting
enough to render it useful to explore the application of the FST convection
in different parts of the HR diagram. We give here one positive example (the
interpretation of the B\"ohm--Vitense gap at B--V$\sim 0.35$) and one
``negative" example, namely the impossibility of explaining the
patterns of Lithium depletion in the Sun and in open clusters with the FST
model. This latter result, however, may be telling us something else on the
behaviour of convection in young convective stars.

\begin{figure}[!ht]
  \begin{center}
    \epsfig{file= 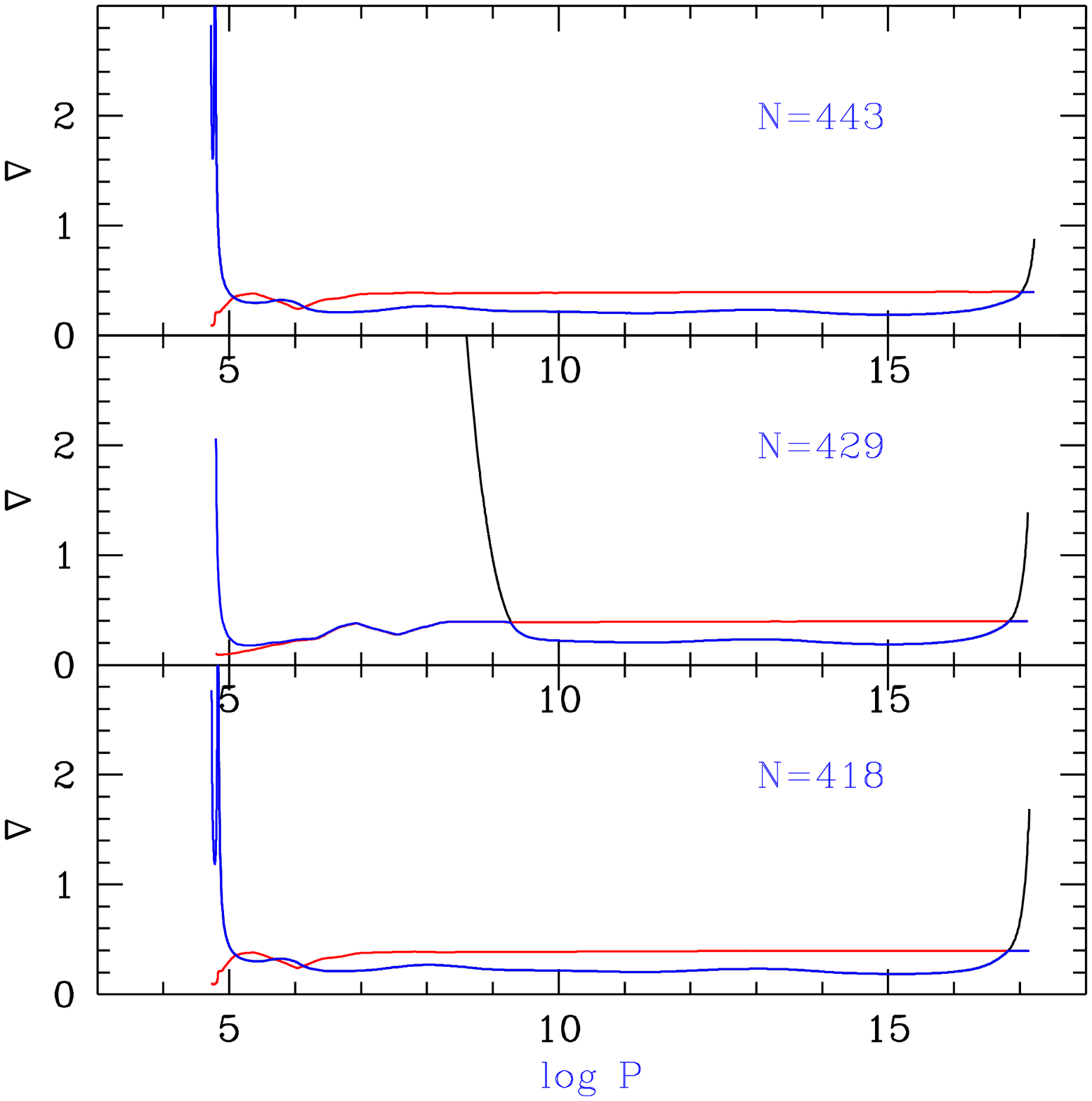, width=7cm}
  \end{center}
\caption{The radiative (black), adiabatic (red) and
superadiabatic (blue) gradient $\nabla-\nabla_{\rm ad}$ in the external
layers of the selected models along the 1.5\msun\ evolution (shown as open
squares in Figure \ref{fig6}, versus the logarithm of the pressure. The first
and third model have their convection region limited to the atmospheric
layers, while the second one shows an extension of convection down to log P
$\sim 9$, well inside the envelope. The transition occurs for a very small
variation of the physical parameters, and is mainly due to the high
inefficiency of the FST model in the atmospheric regions.\label{fig1}}
\end{figure}

\section{The B\"ohm Vitense gap at B--V$\sim 0.35$}

One of the most interesting results in our recent exploration of ``full" FST
models regards the main sequence: the FST convection, due to its very
low efficiency close to the stellar surface, and high efficiency in
the inner subatmospheres, yields a {\it very sharp} transition between
structures which are convective only in the surface layers, and structures
which show a well developped convection also in the interior
(\cite{dantona2002}) .
This is shown in Figure \ref{fig3} by comparing the stratification of models
of different mass in the P--T plane: models down to 1.42\msun\ are mostly
radiative (or convection is so inefficient that the convective gradient
sticks very close to the radiative one), while suddenly, at 1.41\msun, an
extended adiabatic convection region appears for a larger part of the
envelope. The fast increase in the convective mass fraction ($M_{\rm
ce}$) as a function of the main sequence color B--V is shown in Figure 3. 
This characteristic of the models is reflected in their
mass--\teff\ relation, which becomes suddenly steeper around \teff$\simeq
6800$K, and is also apparent in the HR diagram as a sudden change of slope
(Figure 3). Using numerical simulations we have shown
that this feature produces a stellar depletion which is
consistent with the gap seen in the Hyades at 0.33 \simlt B--V \simlt 0.38,
one of the so called ``B\"ohm Vitense gaps" after \cite*{bvcanterna1974} and
\cite*{bv1982}, found by \cite*{rachford-canterna2000} in 6 our of 9 open
clusters which have been investigated. The standard MLT models do not show
this behavior (see Figure \ref{fig5}).

The {\it very sharp} variation of the stellar structures in the HR diagram
(or in the \teff\ gravity plane) is shown as a {\it transition line} in the
HR diagram of Figure \ref{fig6}. Models on the right of this line have
extended convective regions, and models on the left have very small
convective regions, independently from their evolutionary phase (pre, on or
post the main sequence). This can be seen in Figure \ref{fig1}. The
transition line is compared in Figure \ref{fig6} with the location of the
$\delta$ Scuti and $\gamma$ Doradus  instability strips. We may speculate
that our transition line separates HR diagram regions which harbor different
modalities of stellar oscillation patterns, as the excitation or
driving mechanisms can be very different for stars having so different
convective structure. In particular, it could represent the dividing line
between coherent pulsations and solar-type oscillations. In this case, we
should have expected that also the stars between the red edge of the $\gamma$
Doradus instability strip and the transition line are pulsating. Notice that
the location of the red edge of the $\gamma$ Doradus strip is based on
observations from the ground and might still be uncertain in the theoretical
HR diagram. On the other hand, many other structural parameters may have a
role in defining these instability strips, and at least two of these
---elements diffusion and rotation--- should be considered. A further note of
caution on the naivety of this proposal comes also from the complex behavior
of the chromospheric and transition layer indicators for the MS stars on the
right of the transition line (\cite{bv2002}).

\begin{figure}[!ht]
  \begin{center}
    \epsfig{file= 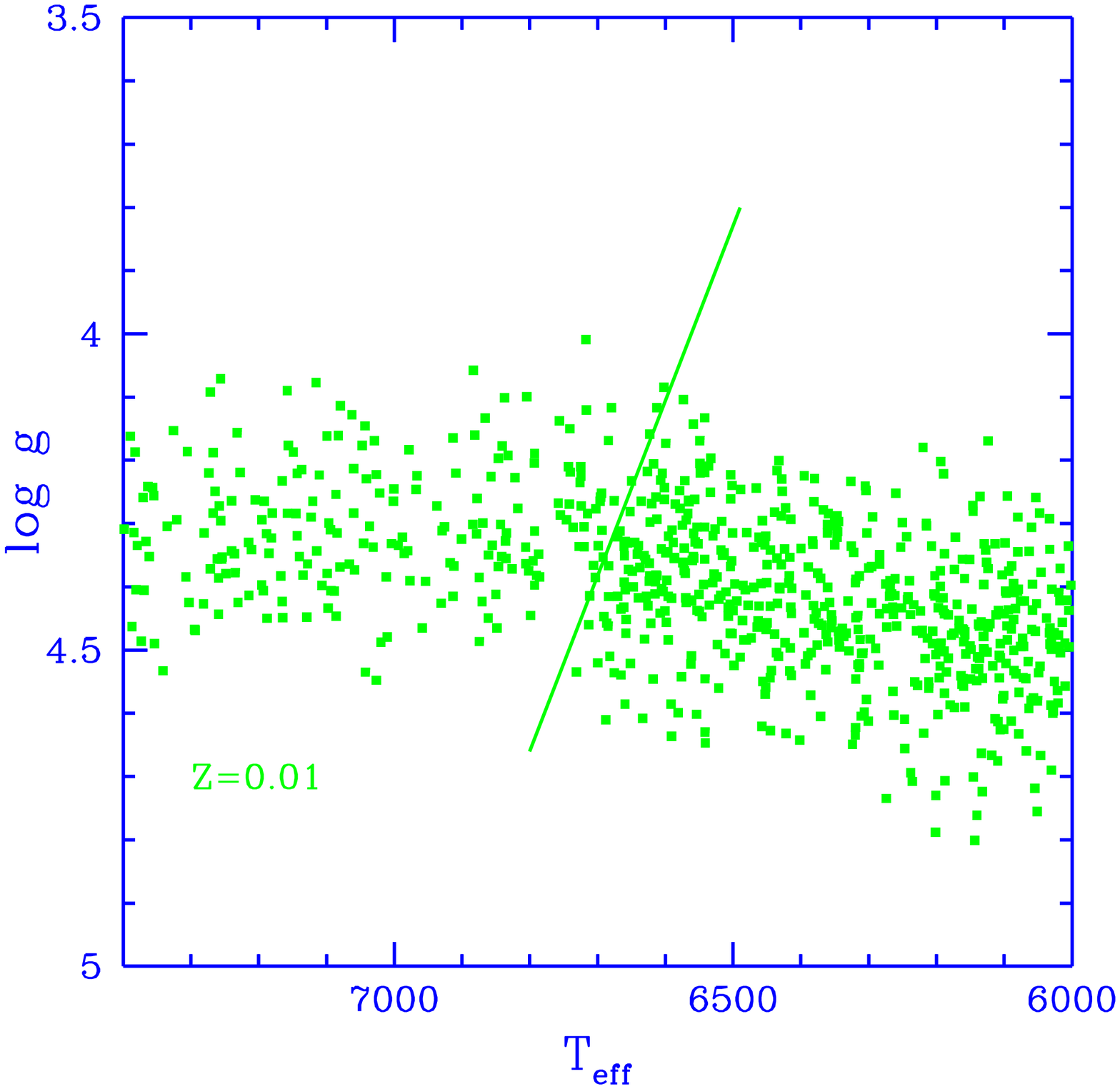,
width=6.8cm}
    \epsfig{file= 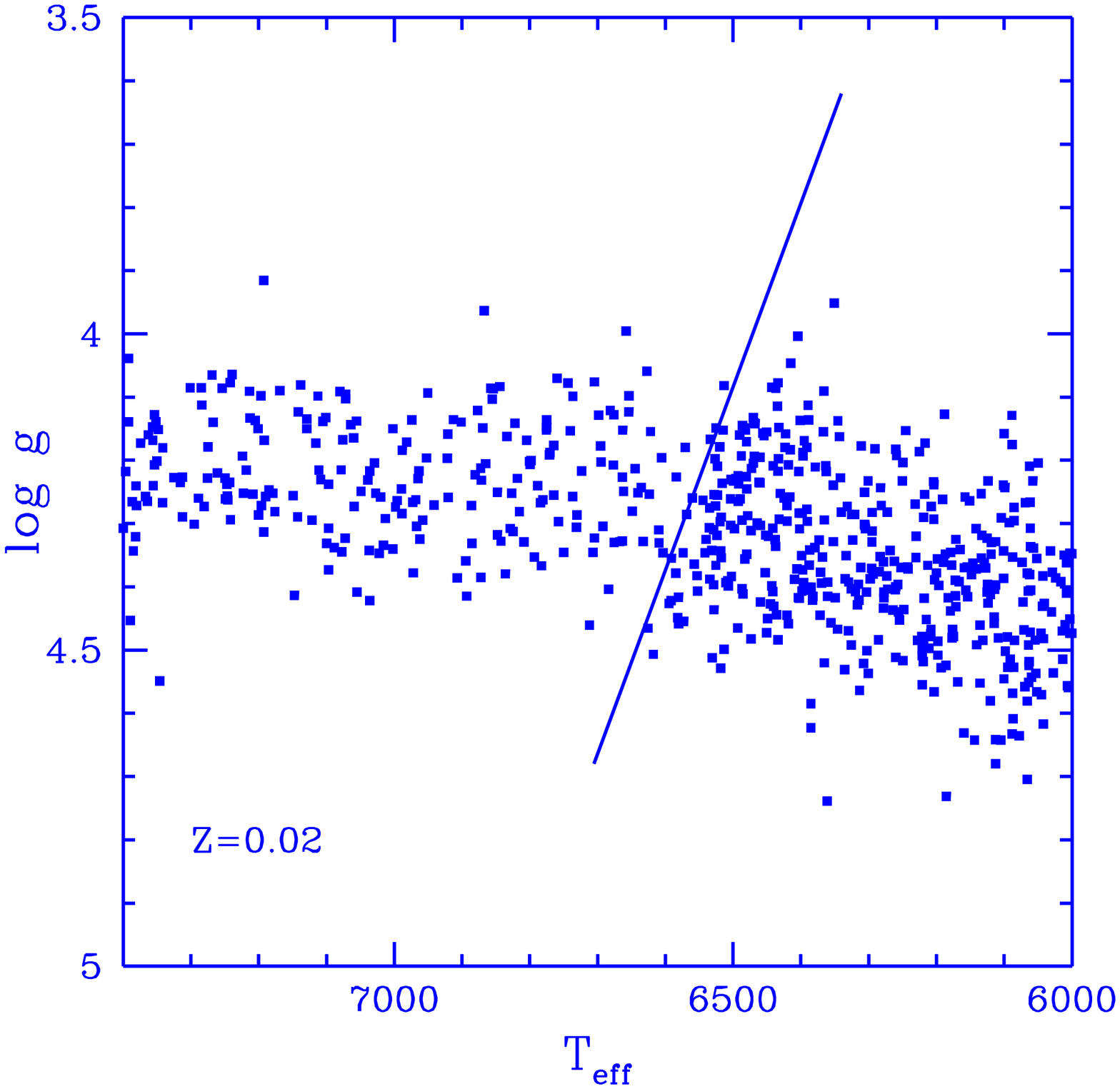,
width=6.8cm}
    \epsfig{file= 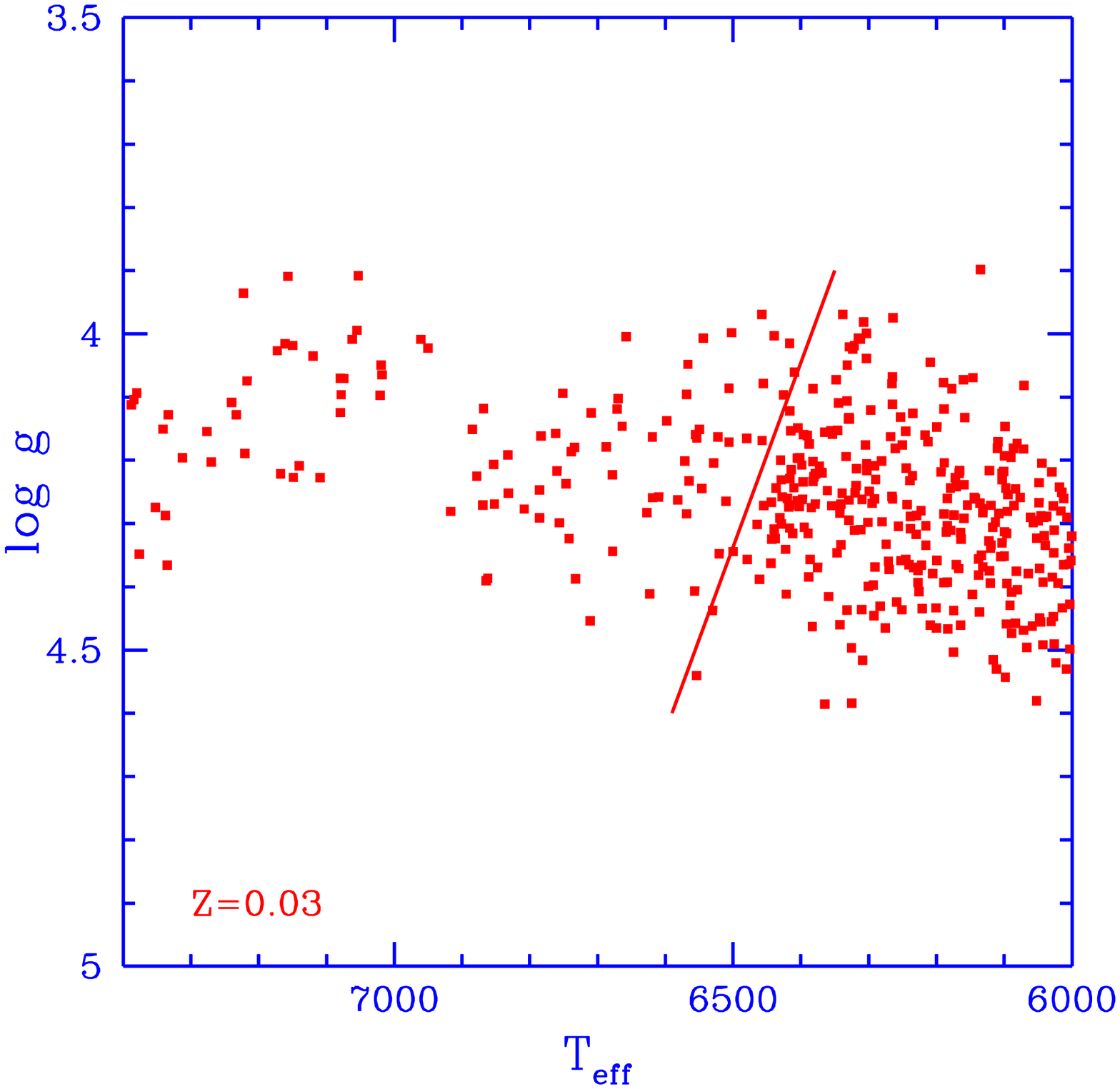,
width=6.8cm}   \end{center}
\caption{Simulations based on the Montalb\'an et al. (2003) models of
different metallicites. A population of 1500 stars is randomly chosen
between ages of $10^7$ and $10^9$yr defines the B\"ohm Vitense gap at
the boundaries indicated by the diagonal lines.\label{fig7}} \end{figure}

Will the B\"ohm Vitense gap appear also in the field stars? Here the problem
will be much more complicated by the blurring due to the fact that we have a
mixture of ages and metallicities! Figure \ref{fig7} shows simulations for
three stellar populations with different metallicities (Z=0.01, 0.02 and
0.03), covering ages from 10$^7$\ to 10$^9$yr, and distributed according to
a Salpeter mass function. The diagonal lines separate the location of the
models having deep envelope convection (at cooler \teff) from those, hotter,
having very thin atmospheric convection. A gap is apparent on the left of
each line, which shift to cooler \teff\ with increasing metallicity, so that
it will not be easy to locate a gap in a sample of stars not homogeneous in
metallicity.

\section{The `historical' problem of Lithium}

In the years 1965-1990, we called `problem of lithium in the Sun'
the inhability of the solar mass evolutionary tracks to burn a substantial
fraction of their initial lithium during the Pre Main Sequence (PMS). This
result was generally taken as a good proof that additional mechanisms for
depletion were required, acting during the long solar MS lifetime, to reduce
by a factor $\sim$140 the initial solar system abundance
($\log$ N(Li)=3.31$\pm 0.04$ (\cite{anders-grevesse1989}).
This interpretation still today is taken as most plausible one, confirmed by
the scarce lithium depletion, at the solar mass, in young open clusters (see
e.g. \cite{chaboyer1998}). In fact the lithium vs. \teff\ relation for the MS
stars of young open clusters indicates a lithium depletion by at most a
factor two for the solar mass in young clusters, while it is compatible with
the solar depletion in some stars of the cluster M67, close to the solar age. For
recent reviews see e.g. \cite*{jeffries2000} and \cite*{pasquini2000}.

However, a different problem emerges from the most recent computation of
solar models: they deplete too much lithium during the PMS evolution
(\cite{dm1994}, \cite{dm1997}, \cite{schlattlweiss1999}, \cite{piau2002}) and
are incompatible with the young open clusters observations. This problem is
most severe in models using very efficient convection models, in fact  it is
more relevant for the D'Antona and Mazzitelli (1994 and 1997) models adopting
the FST convection. MLT models of the most recent  generation, adopting
updated equations of state and opacities also deplete too much lithium.
\cite*{dm2003} have recently reappreciated that the problem is severe {\it
in tracks whose convection is adjusted to provide the solar radius at the
solar age!} If one does not require the solar fit, it is easy to decrease the
convection efficiency and to obtain a negligible PMS lithium depletion.
\begin{figure}[ht]
  \begin{center}
    \epsfig{file= 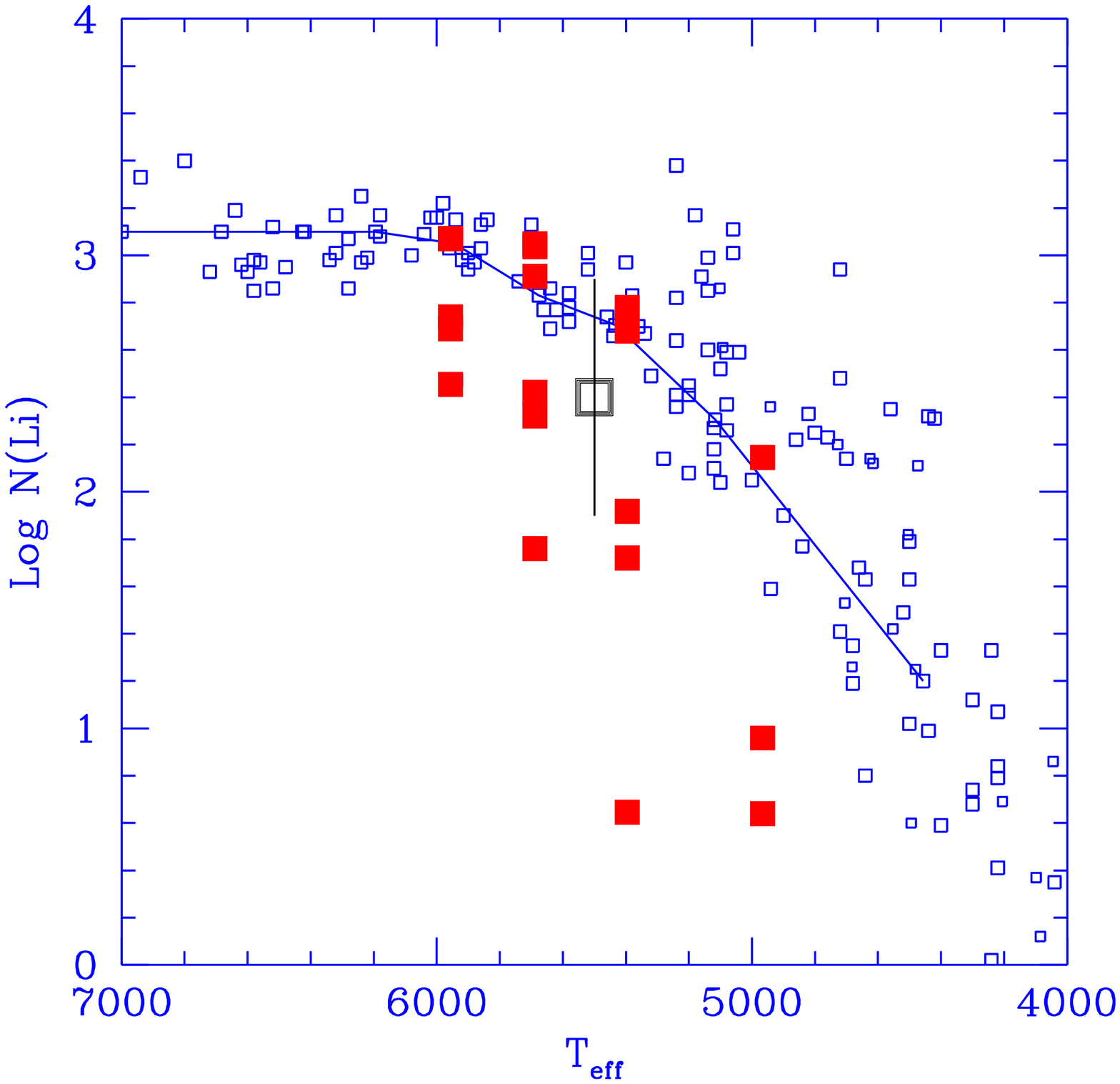, width=7cm}
  \end{center}
\caption{The Pleiades data by Soderblom et al. (1993) and Garcia Lopez et al.
(1994) (open squares) are compared with the depletion predicted by the models
in Table 2 (full big squares). The models are placed at the \teff\ they would
take in an empirical MS, at the Pleiades age. Only the upper squares,
corresponding to the models with  $\alpha_{\rm atm}=\alpha_{\rm in}=1$, are
compatible with the data. The full line shows the depletion from the  models
by Ventura et al. (1998b) computed including the thermal effect of a magnetic
field on the convective temperature gradients. The large open square with the
error bar represents the lithium abundance of the secondary component of
RXJ~0529.4+0041  (log N(Li)=2.4$\pm$0.5, Covino et al. 2001). The \teff\ at
which the point is located (5500K) is assumed to be the main sequence \teff\
of a star of mass 0.925\msun.\label{fig8}}
\end{figure}
\begin{figure}[bt]
  \begin{center}
    \epsfig{file= 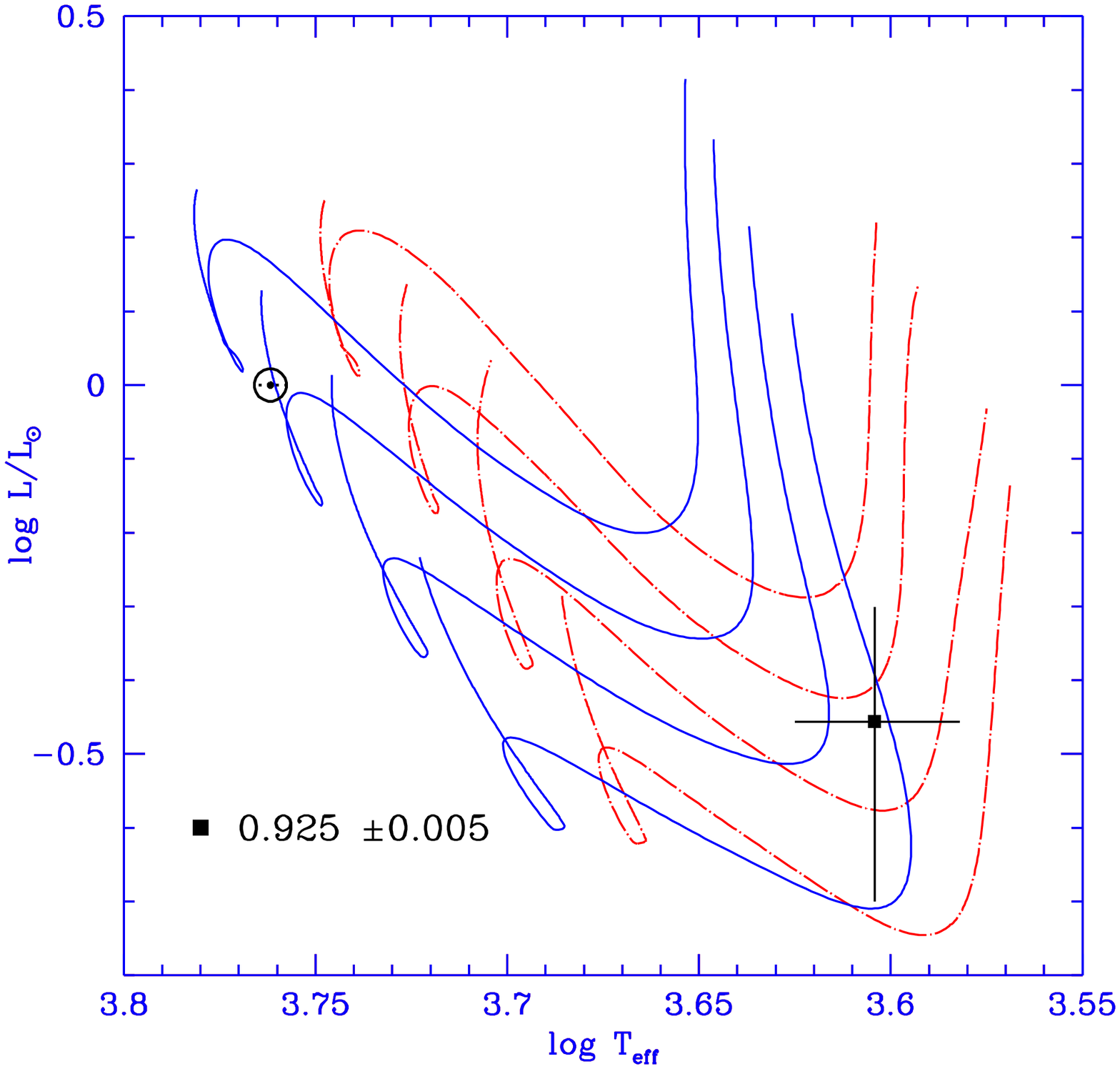, width=7cm}
  \end{center}
\caption{The figure shows the tracks of 0.8, 0.9, 1.0 and 1.1\msun\ computed
with AH97 atmospheres and $\alpha _{\rm in}=1.9$ (full lines, left), or
$\alpha _{\rm in}=1.0$ (dash-dotted lines, right). The solar location is
shown: it is compatible with the solar model having $\alpha _{\rm in}=1.9$,
but $\sim$400K {\it hotter} than the solar model with $\alpha _{\rm in}=1.0$.
The location of the secondary component of the binary RXJ~0529.4+0041 is also
shown. Its mass is 0.925$\pm$0.005\msun, so it is best compatible with the
$\alpha=1$\ models\label{fig9}} \end{figure}
In Fig. \ref{fig8} we compare the lithium depletion predicted by many of our
models, computed using AH97 atmospheres ($\alpha_{\rm atm}=1$) as boundary
conditions, with the Pleiades data by \cite*{soderblom1993} and
\cite*{garcialopez1994}.
Only the upper squares, corresponding to the models
employing $\alpha_{\rm in}=1$, which {\it do not fit the Sun}, are
compatible  with the data. The HR diagram of the models is shown in Figure
\ref{fig9}. The $\alpha=1$\ tracks are on the right, much cooler than what is
needed to allow the solar fit!
In addition, Figure \ref{fig9} opens up the complex problem of the location
in the HR diagram of PMS tracks (for a review see, e.g. \cite{dantona2000},
while our full recent computation for this phase can be found in
\cite{mdkh2003}). When we compare the location of PMS theoretical tracks with
the observed few data of PMS stars for which an independent determination of
mass is available\footnote{these stars either belong to binaries (e.g.
\cite{covino2001}, \cite{steffen2001}, or one can measure the stellar mass
from the dynamical properties of their  protoplanetary
disks (\cite{simon2000}).}, the tracks most consistent with the observations
are again those with {\it cooler atmospheres} (higher mass for a given
spectral type) and thus those which, generally, provide a radius larger than
R$_\odot$\ for the solar model. In fact, Figure \ref{fig9} shows that the
$\alpha=1$\ tracks are well compatible with the location of the secondary
component of the eclipsing spectroscopic binary RXJ 0529.4+0041, one of the
best determined PMS masses (\cite{covino2001}. It is then clear that {\it
1)} the HR diagram location of the tracks during the PMS evolution, and
{\it 2)} PMS lithium depletion, are two problems correlated each other, as it
could have been expected, because, the smaller the \teff\ of the Hayashi
track, the smaller is the temperature at the base of the convective envelope
during the possible phase of lithium burning. Both the HR diagram location
and the lithium depletion seem to be compatible {\it only} with models in
which {\it PMS convection is much less efficient than MS convection}. Is this
simply another proof that we are not able to model convection, or that there
are unsolved problems with the opacities? Or does it mean that there is some
other parameter playing a role in the PMS --and not on the MS? It is probably
too early to derive strong conclusions, but we have suggested that PMS
convection is inhibited by the thermal role of a dynamo built magnetic field
(\cite{ventura1998b}, \cite{dantona2000}).

\section{Non-local convection in the Sun}

Always in the framework of stellar evolution for asteroseismology, we
performed one more consistency test, of a completely different nature. In
this case, our attention is not at the surface, but at the bottom
of the solar convective zone (CZ). In fact, a constraint to the thickness of
the overshooting layer in this region has been set by helioseismology (e.g.
\cite{basu-antia1997}), and it can not be larger than $\sim$0.05$H_p$.
We constructed a detailed FST solar model matching the correct thickness
of the CZ (the surface boundary condictions are in this case of negligible
importance) and applied the treatment suggested by
\cite{canuto-dubovikov1998} (CD98) to a thin region centered around the
formal Schwarzschild boundary of convection, to gain insight on what happens
when a fully non-local turbulence theory is applied to a stellar structure.

\begin{figure}[!ht]
  \begin{center}
    \epsfig{file= 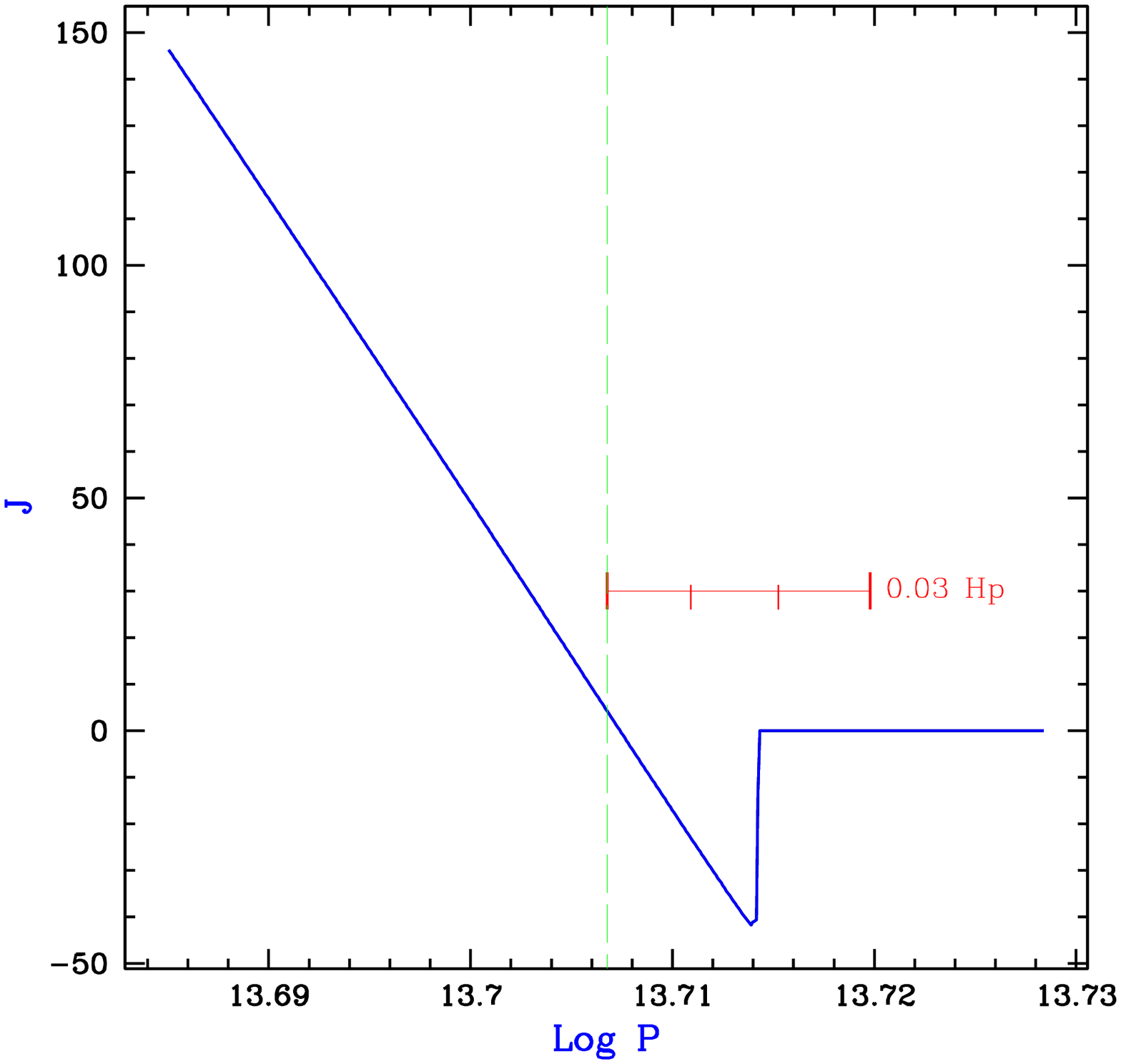, width=7cm}
  \end{center}
\caption{The non-local turbulent flux $J$ (cgs units) around the local
Schwarzschild boundary (vertical dashed line) is shown as a function of
the logarithm of the pressure P (cgs units). The flux is obviously negative
in the overshooting region and, contrarily to what one could expect according
to a thumb rule (perhaps exponential decay), a very sharp decay of the flux
is found, suddenly ending the extra-mixed region less than 0.02$H_p$\ below
the local boundary.\label{jei}} \end{figure}

\begin{figure}[bt]
  \begin{center}
    \epsfig{file= 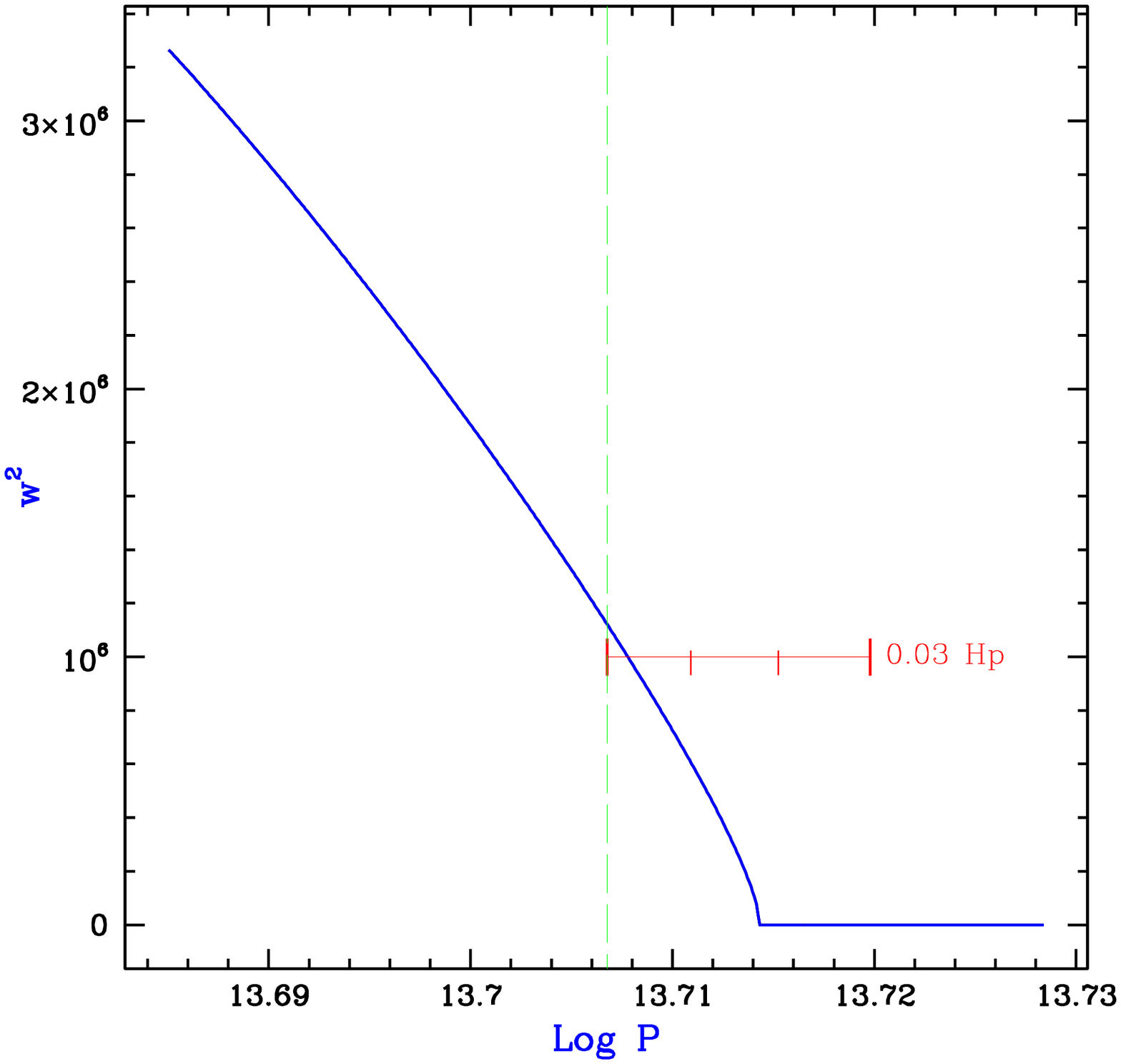, width=7cm}
  \end{center}
\caption{The mean quadratic velocity variance $\overline{w}^2$. Velocity
decays outside the local Schwarzschild boundary with the same slope as
inside. When it approaches zero, however, the slope becomes steeper, and
overshooting comes to a full stop. This suggests that all the
evolutionary computations performed up today in the hypotesis of an
exponential decay of the diffusive coefficient outside formally
convective regions can perhaps (largely?) overestimate the amount of
extra-mixed matter.
\label{wm2}} \end{figure}

More in detail, we have computed from the local model the starting
distribution for the quantities: $K$ (turbulent kinetic energy in the radial
direction), $\overline\theta^2$ (mean qua\-dra\-tic temperature variance),
$J$ (convective flux), $\overline{w}^2$ (mean qua\-dra\-tic velocity
variance) and $\epsilon$ (dissipation rate), according to (42a-c), (43a-c)
and (44a-d) in CD98. Then, temporal relaxation to the five above quantities
has been allowed, according to the equations (19a-d) and (35a-b), until
stationary conditions were reached. For each relaxation step, the gradient
$\beta$:

\begin{equation}
    \beta = { T \over {H_p}}(\nabla - \nabla_{ad})
\end{equation}

has been updated according to:
\begin{equation}
    \beta = \beta_{rad} - { {J+F(K)} \over \chi}
\end{equation}
where $\chi$ is the thermometric conductivity, $\beta_{rad}=T(\nabla_{rad} -
\nabla_{ad})/H_p$. Further, $F(K)$\ is the kinetic energy flow $-{\nu_t \over
{2c_p}}{\partial K \over \partial r}$ and  $\nu_t$\ is the turbulent
viscosity, $\nu_t \propto {K^2 \over \epsilon}$.

Eq. (18c) from CD98, relative to the temperature, was not included in
the final network since, being superadiabaticity at the bottom of the
solar CZ negligible, temperature itself turned out to be nearly
stationary already at the beginning of the relaxation.

As for the diffusive terms $D_f$, the most simple down-gradient
approximation has been chosen, namely, for each generic turbulent quantity
$Q$:
\begin{equation}
D_f(Q) \propto \nu_t {\partial Q \over \partial r}
\end{equation}
This is (perhaps) far from the best one can do but, noticeably, no
built-in scale length is present, contrarily to the only other non-local
treatment (\cite{xiong1985}) with which extensive stellar models have been
computed. In fact, in Xiong case, an explicit, arbitrary scale length
was included, and the results almost linearly depended on the value of
the scale length.
After $2-3 \times 10^6$ s, all the six quantities (including $\beta$)
reached a final, stationary distribution, clearly showing a thin overshooting
region.
Figure \ref{jei} shows the behavior of the turbulent flux $J$; Figure
\ref{wm2} presents the mean quadratic velocity variance $\overline{w}^2$.
Overshooting is present indeed, but its amplitude does not overcome
0.02$H_p$, consistently with the solar observational
constraints.\footnote{The very sharp decline of the flux is an important
constraint for the `local' parametrization of overshooting. In particular, it
is different from the approximation of exponential decay adopted, e.g., by
\cite{ventura1998a} and \cite{herwig}. These results apply only to
overshooting below the CZ, we can not extend them to what happens above the
convective cores.} The same relaxations have been performed for the CZ of the
solar mass track at various evolutionary phases, from the first appearence of
a radiative nucleus in PMS, up to early red giant. The overall features turn
out to be similar in all cases. It is found that the overshooting region is
absolutely negligible ($<0.005 H_p$) in PMS during the lithium burning phase.
This at least ensures us that inclusion of proper overshooting in PMS
evolutionary models will not worsen the problems with the exceedingly large
solar lithium depletion shown by todays standard models. The thickness of
overshooting, then, steadily increases during the evolution, reaching $\sim
0.06H_p$\ at $\log L/L_\odot =1$, where the computations have been stopped.
In all case, the decay of $\overline{w}^2$\ is very sharp, putting an end to
overshooting.

The only conclusion we can presently draw from these first non-local
results is that probably the chosen approximation for the diffusive
terms is perhaps not too bad, at least as long as the thickness of
overshooting is concerned, since they are consistent with the
observational solar constraints, and do not worsen the lithium problem.

\begin{acknowledgements}

  This work has been partially supported by the MIUR through the COFIN
  2002-2003 project ``Asteroseismology'.

\end{acknowledgements}

\end{document}